\documentclass[pdflatex,sn-mathphys-num]{sn-jnl}

\usepackage{graphicx}%
\usepackage{multirow}%
\usepackage{amsmath,amssymb,amsfonts}%
\usepackage{amsthm}%
\usepackage{mathrsfs}%
\usepackage[title]{appendix}%
\usepackage{xcolor}%
\usepackage{textcomp}%
\usepackage{manyfoot}%
\usepackage{booktabs}%
\usepackage{braket}
\usepackage{algorithm}%
\usepackage{algorithmicx}%
\usepackage{algpseudocode}%
\usepackage{listings}%
\usepackage{bm}
\usepackage{xcolor}

\theoremstyle{thmstyleone}%
\theoremstyle{thmstyletwo}%

\theoremstyle{thmstylethree}%

\begin{document}

\title[Article Title]{Experimental simulation of non-equilibrium quantum piston on a programmable photonic quantum computer}


\author*[1]{\fnm{Govind} \sur{Krishna}}\email{govindk@kth.se}

\author[1]{\fnm{Rohan} \sur{Yadgirkar}}

\author[1]{\fnm{Balakrishnan} \sur{Krishnakumar}}

\author[1]{\fnm{Andrea} \sur{Cataldo}}

\author[1]{\fnm{Ze-Sheng} \sur{Xu}}

\author[2]{\fnm{Johannes W. N.} \sur{Los}}

\author[3]{\fnm{Val} \sur{Zwiller}}

\author*[1,4]{\fnm{Jun} \sur{Gao}}\email{jungao@hust.edu.cn}

\author*[1]{\fnm{Ali} \sur{W.} \sur{Elshaari}}\email{elshaari@kth.se}

\affil[1]{\orgdiv{Department of Applied Physics}, \orgname{KTH Royal Institute of Technology}, \orgaddress{\street{Albanova University Centre, Roslagstullsbacken 21}, \city{Stockholm}, \postcode{106 91}, \state{Stockholm}, \country{Sweden}}}

\affil[2]{\orgname{Single Quantum B.V.}, \orgaddress{\city{Delft}, \postcode{2628 CJ}, \country{The Netherlands}}}

\affil[3]{\orgname{RISE Research Institutes of Sweden}, \orgaddress{\city{Stockholm}, \country{Sweden}}}

\affil[4]{\orgdiv{School of Optical and Electronic Information}, \orgname{Huazhong University of Science and Technology}, \orgaddress{\street{Luoyu Road 1037}, \city{Wuhan}, \postcode{430074}, \state{Hubei}, \country{China}}}


\abstract{
Quantum fluctuation relations provide a microscopic formulation of thermodynamics beyond equilibrium, but experimentally accessing many-body quantum work statistics remains an outstanding challenge. The quantum piston constitutes a canonical model of boundary-driven nonequilibrium dynamics, where finite-time deformation of a confining potential generates non-adiabatic transitions, dissipation and irreversibility. Here we experimentally simulate the nonequilibrium dynamics of a two-boson quantum piston on a programmable photonic quantum computer. Using two indistinguishable photons, we encode a truncated piston propagator through a quasi-unitary embedding, with an ancilla mode representing leakage into higher-energy states outside the resolved manifold. This architecture enables direct reconstruction of thermodynamic transition statistics for both expansion and compression protocols as functions of driving speed and final trap length. We observe the crossover from quasi-adiabatic to strongly non-adiabatic evolution and show that bosonic interference restructures the resulting two-particle Fock-state populations and work distributions. The measured statistics are in close agreement with theoretical predictions and satisfy the Jarzynski equality across expansion and compression protocols for cyclic driving we further quantify irreversibility through dissipated work and state overlap. Our work identifies programmable photonic quantum hardware as a powerful platform for simulating nonequilibrium quantum thermodynamics and for experimentally resolving how indistinguishability and many-body interference shape quantum work, dissipation and entropy production.
}




\maketitle

Understanding how energy is exchanged in nonequilibrium processes lies at the heart of thermodynamics and has thus attracted extensive research interests.~\cite{Jarzynski1997, Crooks1999, Quan2012, Hatano2001} While classical thermodynamics primarily describes average quantities such as work and heat, modern fluctuation relations reveal that microscopic systems exhibit full probability distributions of work, even far from equilibrium. The Jarzynski equality and Crooks fluctuation theorem establish that these distributions encode equilibrium free-energy differences, thereby linking irreversibility observed at the macroscopic level to the microscopic reversibility of the underlying dynamics.~\cite{Jarzynski1997, Crooks1999, Quan2012} 
In the quantum regime, however, work is not associated with a Hermitian operator like familiar observables such as position or energy. Instead, it is defined operationally through a two-time projective energy-measurement protocol: the system's energy is measured once before driving begins and once after it ends, with work identified as the difference between these outcomes. This yields a work distribution fully determined by transition probabilities between instantaneous energy eigenstates.~\cite{Talkner2007,Campisi2011, Kafri2012}

\begin{figure}[t]
  \centering
  \includegraphics[width=\linewidth]{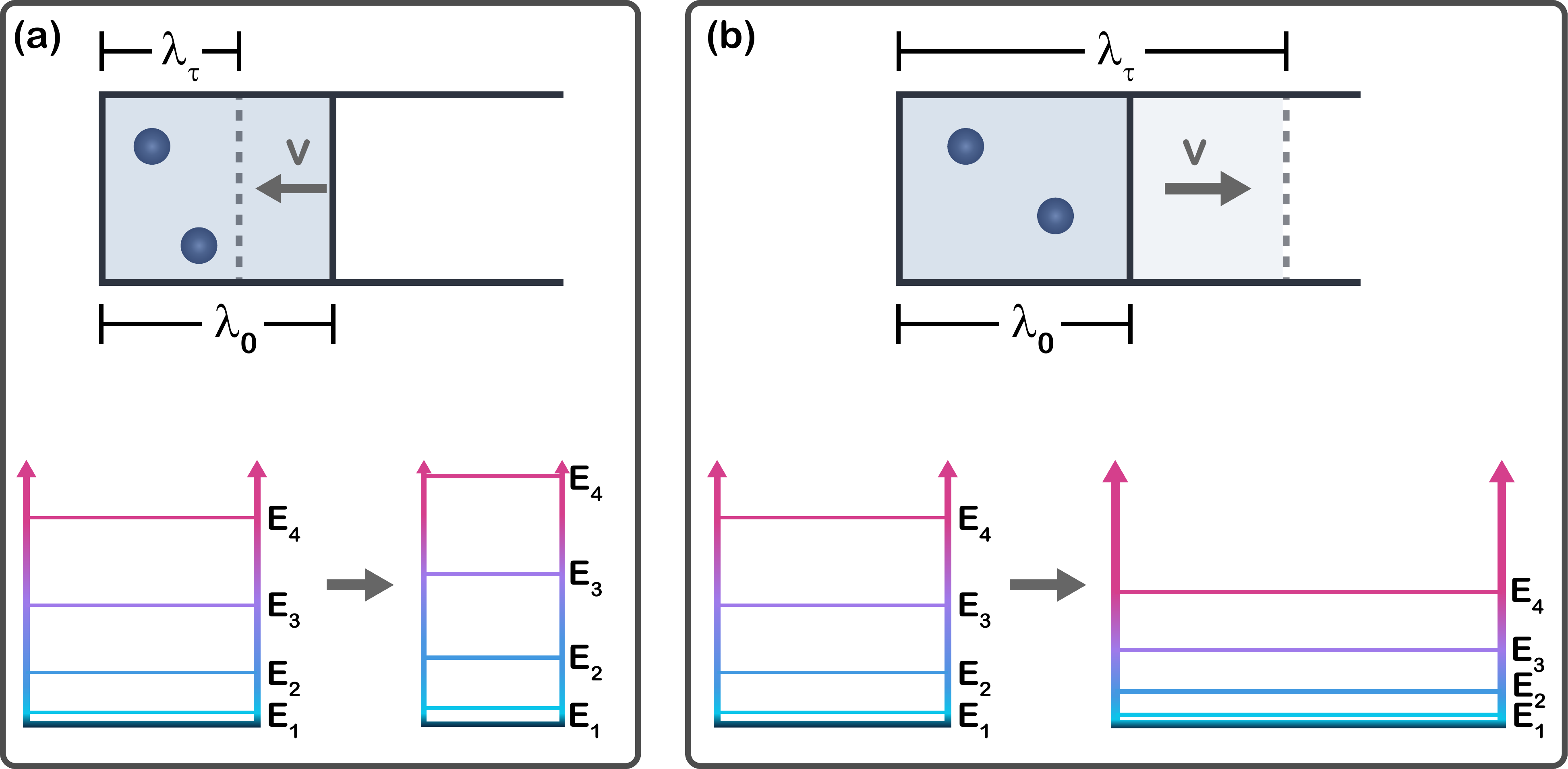}
 \caption{\textbf{Quantum piston expansion and compression protocol.}
Identical bosons of mass $M$ are confined in a one-dimensional box of time-dependent length $\lambda(t)$. At $t=0$, the system is prepared in a thermal quantum state at temperature $T$, corresponding to a Gibbs mixture of energy eigenstates $\{|i^{\lambda_0}\rangle\}$ with Boltzmann weights. The single-particle energy levels are $E_n(\lambda)=\frac{\hbar^2\pi^2 n^2}{2m\lambda^2}, \quad n=1,2,\dots$. Then the system is detached from the bath and one boundary moves from $\lambda_0$ to $\lambda_\tau$ during time $\tau$ with constant velocity $v=(\lambda_\tau-\lambda_0)/\tau$. Expansion ($\lambda_\tau>\lambda_0$) decreases the level spacing, while compression ($\lambda_\tau<\lambda_0$) increases it. The resulting unitary evolution induces transitions between initial eigenstates $\{|i^{\lambda_0}\rangle\}$ and final eigenstates $\{|f^{\lambda_\tau}\rangle\}$. Within the two-projective-measurement scheme, quantum work is defined as $W=E_f(\lambda_\tau)-E_i(\lambda_0)$, and its distribution is determined by the thermal occupation probabilities and the transition amplitudes generated by the piston evolution.}
  \label{fig:quantum_piston_schematic}
\end{figure}

A paradigmatic model for studying nonequilibrium quantum work is the quantum piston, in which a particle is confined in a one-dimensional box whose boundary is dynamically displaced in time.~\cite{Zhu2016, Quan2007, Deffner2008} Changing the piston length modifies the quantized energy spectrum (Fig.~\ref{fig:quantum_piston_schematic}) and, during a finite-time driving protocol, induces transitions between energy levels whose probabilities depend on the driving speed and trajectory. In the adiabatic limit, where the boundary moves slowly compared with the internal timescales of the system, populations remain in the instantaneous eigenbasis and the work distribution is narrow and nearly reversible. In contrast, rapid driving activates strong non-adiabatic transitions, broadening the distribution and enhancing entropy production and irreversibility.~\cite{Jarzynski1997}
Because of this clear and controllable crossover between regimes, the quantum piston has become a fundamental testbed for exploring the microscopic structure of quantum work statistics.~\cite{Campisi2011}

Extending this framework to multiple identical particles introduces fundamentally new physics that has no classical analogue.
Transition probabilities between many-body eigenstates are then constrained by quantum statistics and, in general, cannot be written as simple products of single-particle amplitudes.~\cite{Gong2014}
For non-interacting bosons, multi-particle transition amplitudes are given by matrix permanents of the single-particle transition-amplitude matrix, giving rise to characteristic interference effects that are absent for distinguishable particles.~\cite{Gong2014,Tichy2014}
Recent theoretical work has shown that these multi-boson quantum piston work distributions can be efficiently simulated using linear-optical interferometers, establishing a direct mapping between nonequilibrium quantum thermodynamics and boson-sampling architectures based on photons and beam splitters.~\cite{Liu2024,Aaronson2011}

Here we experimentally realize this proposal by implementing a programmable quantum-piston transformation in a universal integrated photonic circuit.
Using two indistinguishable photons propagating through a $12\times 12$ Clements interferometer,~\cite{Clements2016}
we encode a truncated four-level piston propagator via a quasi-unitary extension into a five-mode transformation~\cite{Tischler2018, krishna2025emulationcoherentabsorptionquantum}, where a single ancilla mode represents population leaking into higher-energy states beyond the encoded subspace.~\cite{Liu2024}
By systematically varying the effective piston speed and final length, we directly observe how bosonic interference reshapes the two-photon Fock-state distribution and the resulting quantum work statistics.
Our results demonstrate controlled simulation of nonequilibrium quantum thermodynamics in a programmable photonic platform and reveal how many-body interference governs the structure and robustness of quantum work distributions.~\cite{Gong2014,Liu2024}

\section*{Results}

\subsection*{Quantum piston formalism and transformation matrix synthesis}

The quantum piston describes identical bosons of mass $M$ confined in a one-dimensional box whose length changes from $\lambda_0$ to $\lambda_\tau$ during protocol time $\tau$ (Fig.~\ref{fig:quantum_piston_schematic}). The instantaneous single-particle eigenenergies are

\begin{equation}
E_n(\lambda)=\frac{\hbar^2\pi^2 n^2}{2m\lambda^2}, \qquad n=1,2,\dots
\end{equation}

For $N$ non-interacting identical bosons the many-body eigenstates are bosonic Fock states $|\mathbf{n}^\lambda\rangle = |n_1,n_2,n_3,\dots\rangle$ with $\sum_k n_k=N$ and total energy $E_{\mathbf{n}}(\lambda)=\sum_k n_k E_k(\lambda)$. At $t=0$ the system is prepared in a thermal Gibbs state

\begin{equation}
\rho_0=\sum_{\mathbf{n}} p_{\mathbf{n}}\,|\mathbf{n}^{\lambda_0}\rangle
\langle\mathbf{n}^{\lambda_0}|, \qquad
p_{\mathbf{n}}=\frac{e^{-\beta E_{\mathbf{n}}(\lambda_0)}}{Z_0}, \qquad
Z_0=\sum_{\mathbf{n}} e^{-\beta E_{\mathbf{n}}(\lambda_0)},
\end{equation}

where $\beta=1/k_BT$ and the sums run over all $N$-boson Fock states. The piston motion generates unitary evolution $\hat{U}(\tau)$, inducing transitions between initial and final eigenstates with amplitudes

\begin{equation}
T_{\mathbf{f}\mathbf{i}}(v,\lambda_0,\lambda_\tau)
=\langle\mathbf{f}^{\lambda_\tau}|\hat{U}(\tau)|\mathbf{i}^{\lambda_0}\rangle,
\end{equation}

whose analytical expressions and numerical evaluation are detailed in the Methods section. Within the two-projective-measurement scheme, quantum work is defined as

\begin{equation}
W = E_{\mathbf{f}}(\lambda_\tau) - E_{\mathbf{i}}(\lambda_0).
\end{equation}

For experimental implementation we truncate to the lowest four single-particle levels, defining the transformation matrix $T_{4\times4}(v,\lambda_0,\lambda_\tau)$ over this subspace. Population transferred outside it is unrepresented by this matrix, so $T_{4\times4}^\dagger T_{4\times4}\neq I_4$, reflecting physical leakage into higher levels. Since passive linear-optical interferometers require unitary transformations, $T_{4\times4}$ is embedded into a larger unitary $U_{5\times5}$ by introducing a single ancilla mode representing the collective higher-level subspace~\cite{Tischler2018}. Restricting to one ancilla keeps the implementation compact while providing sufficient accuracy; further details of the quasi-unitary extension and the associated unitarity error are provided in the Methods section.

\begin{figure}[t]
  \centering
  \includegraphics[width=\linewidth]{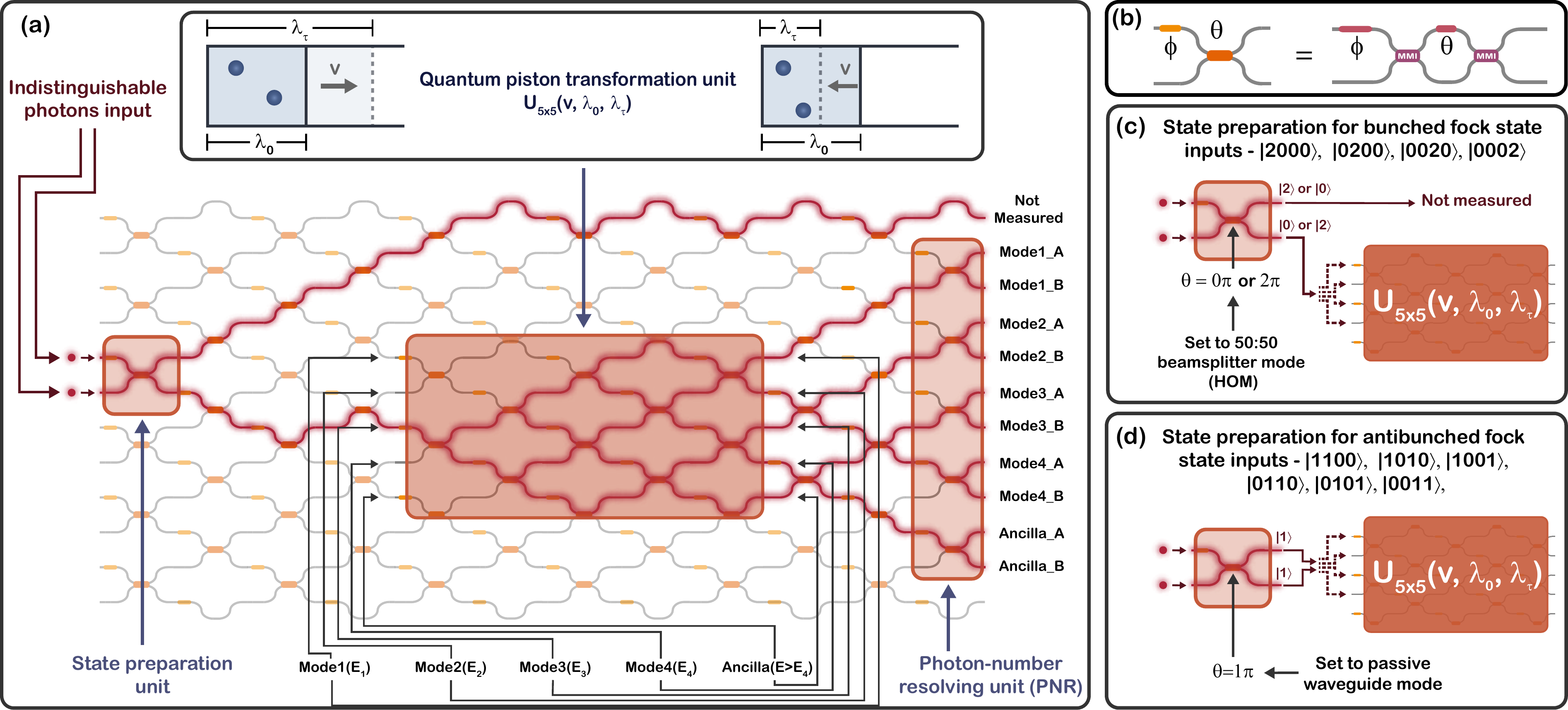}
  \caption{\textbf{Programmable photonic implementation of the quantum piston transformation.}
  (\textbf{a}) Schematic of the universal $12\times12$ Clements interferometer used to implement the quantum piston transformation. The chip is partitioned into a state-preparation unit (left), a programmable
  quantum piston transformation unit realizing an effective $U_{5\times5}(v,\lambda_0,\lambda_\tau)$,
  and a photon-number-resolving (PNR) unit (right). The highlighted five-mode region implements a quasi-unitary extension of the truncated $4\times4$ piston matrix. Each spatial mode in this region corresponds to a piston energy eigenstate: the top four physical modes represent the lowest energy levels $\{E_1,E_2,E_3,E_4\}$ (top to bottom),
  while the fifth mode serves as an \emph{ancilla} representing all higher energy levels together ($E>E_4$). Remaining modes of the $12$-mode mesh are configured in passive routing cross ($\theta=0,2\pi$)/bar ($\theta=1\pi$) states. (\textbf{b}) Representative Mach--Zehnder interferometer (MZI). Left: the simplified two-mode element used symbolically in panel (a) for circuit clarity. Right: the physical MZI structure comprising two fixed 50:50 beam splitters and tunable internal ($\theta$) and external ($\phi$) phase shifters, enabling arbitrary SU(2) transformations required by the Clements decomposition. (\textbf{c}) Preparation of bunched two-photon Fock states (e.g., $\ket{2000}$, $\ket{0200}$, $\ket{0020}$, $\ket{0002}$) by configuring the preparation MZI as a balanced 50:50 beam splitter, corresponding to Hong-Ou-Mandel interference. (\textbf{d}) Preparation of antibunched two-photon Fock states (e.g., $\ket{1100}$, $\ket{1010}$, $\ket{1001}$, $\ket{0110}$, $\ket{0101}$, $\ket{0011}$) by routing photons through passive waveguide configurations prior to the piston unit.}
\label{fig:chip_architecture}
\end{figure}

\subsection*{Experimental implementation and measurement protocol}

The transformation $U_{5\times5}(v,\lambda_0,\lambda_\tau)$ is implemented on a programmable photonic quantum processor, which we refer to as \textbf{Noor-Q}, built around a universal $12\times12$ integrated interferometer based on the Clements rectangular architecture~\cite{Clements2016} (Fig.~\ref{fig:chip_architecture}(a)). The circuit is partitioned into a state-preparation unit, a central piston transformation block, and a photon-number-resolving (PNR) stage~\cite{krishna2025emulationcoherentabsorptionquantum}. Each tunable MZI comprises two fixed 50:50 MMI beam splitters and two thermo-optic phase shifters $(\theta,\phi)$; cascading these realizes an arbitrary $N\times N$ unitary via SU(2) rotations (Fig.~\ref{fig:chip_architecture}(b)). Of the 12 spatial modes, only five are programmed: four encoding piston levels $E_1$--$E_4$ and one ancilla, with all remaining MZIs set to passive routing states. For each parameter set $(v,\lambda_0,\lambda_\tau)$ we compute $T_{4\times4}$, construct $U_{5\times5}$, perform the Clements decomposition, and program the resulting phases onto the piston submesh (see Methods).

Work distributions for a thermal initial state are reconstructed via a conditional-measurement strategy: we measure the output statistics for each of the ten two-photon Fock-basis inputs

\begin{equation}
|2000\rangle,\ |1100\rangle,\ |1010\rangle,\ |1001\rangle,\ |0200\rangle,\
|0110\rangle,\ |0101\rangle,\ |0020\rangle,\ |0011\rangle,\ |0002\rangle,
\end{equation}

and combine the resulting conditional distributions $P_{\mathrm{out}|n}$ with thermal weights $w_n$ to obtain

\begin{equation}
P_{\mathrm{out}}=\sum_n w_n\,P_{\mathrm{out}|n},
\end{equation}

where $w_n$ is the Gibbs occupation probability of state $|n\rangle$ at $\lambda_0$ and temperature T. These ten inputs fall into two preparation classes, each requiring a distinct state-preparation configuration (Fig.~\ref{fig:chip_architecture}(c,d)).

\subsection*{State preparation submesh}

The state-preparation unit is a single tunable MZI whose setting selects between two input classes. \emph{Bunched states} $\{|2000\rangle,|0200\rangle,|0020\rangle,|0002\rangle\}$ are prepared with the MZI at $\theta=0.5\pi$ (balanced 50:50), exploiting Hong--Ou--Mandel interference

\begin{equation}
|1,1\rangle \;\rightarrow\; \tfrac{1}{\sqrt{2}}\bigl(|2,0\rangle+|0,2\rangle\bigr);
\end{equation}

one output arm is left unconnected so only $50\%$ of events enter the piston submesh, and integration times are doubled accordingly. \emph{Antibunched states} $\{|1100\rangle,|1010\rangle,|1001\rangle,|0110\rangle,|0101\rangle,|0011\rangle\}$ are prepared with the MZI at $\theta=\pi$ (passive routing), which directs the two photons deterministically into separate output arms:

\begin{equation}
|1,1\rangle \;\rightarrow\; |1,1\rangle.
\end{equation}

\begin{figure*}[t]
  \centering
  \includegraphics[width=\textwidth]{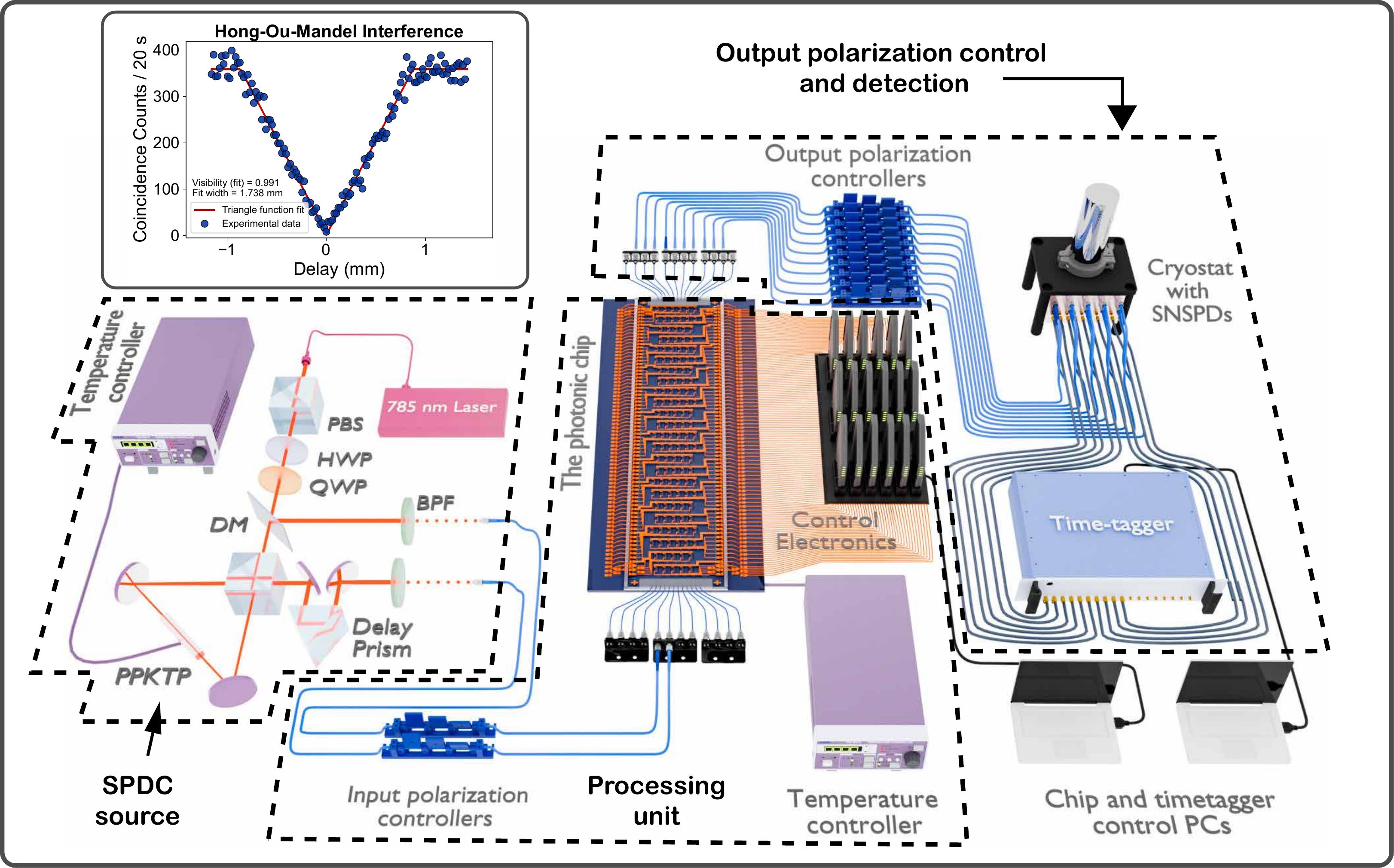}
  \caption{\textbf{Experimental setup for the programmable quantum piston experiment implemented on the photonic quantum processor \textbf{Noor-Q}.}
  Correlated photon pairs are generated via type-II spontaneous parametric down-conversion (SPDC) in a periodically poled KTP (PPKTP) crystal pumped by a 785\,nm continuous-wave laser.
  The degenerate photons (centered near 1570\,nm) are collected into single-mode fibers and injected into a 12$\times$12 universal interferometer based on the Clements architecture.
  Fiber polarization controllers before the chip ensure efficient coupling to the TE mode, and polarization control after the chip maximizes detector efficiency.
  The thermo-optic phase shifters within the interferometer are driven by low-noise current sources with active temperature stabilization.
  All output modes are connected to superconducting nanowire single-photon detectors (SNSPDs) housed in a 2.6\,K cryostat, and coincidence events are recorded using a multi-channel Swabian Time Tagger with a 2\,ns coincidence window. The inset shows the Hong–Ou–Mandel interference measured at the state-preparation stage. The coincidence data are fitted with a triangular function $y = a - b|x - x_0|$, yielding a visibility of 0.991 and a coherence width of 1.738\,mm in optical delay. The delay is tuned by translating one photon path prior to chip injection.
  The high visibility confirms strong photon indistinguishability and temporal overlap, ensuring that the observed two-photon dynamics are governed by bosonic interference in the programmable piston transformation.}
  \label{fig:experimental_setup}
\end{figure*}

\subsection*{Experimental setup}

Fig.~\ref{fig:experimental_setup} summarizes the experimental architecture. Photon pairs from a type-II SPDC source are injected into a universal $12\times12$ Clements interferometer, with all output modes connected to SNSPDs and recorded by a multi-channel
time tagger (see caption and Methods for full source and platform specifications). Photon-number resolution is achieved by routing each of the five piston output modes through a fixed $50{:}50$ MZI, splitting each spatial mode into two detection channels and yielding ten effective PNR channels in total. Two-photon output states are represented in the five-mode Fock basis $|n_a, n_4, n_3, n_2, n_1\rangle$, where $n_a$ denotes the ancilla mode occupation and $n_k$ the occupation of piston level $E_k$. As only eight SNSPDs are available simultaneously, measurements are carried out in three sequential detector configurations, each monitoring a distinct subset of eight channels with the interferometer settings held fixed; the union of coincidence data across all three configurations enables complete reconstruction of the two-photon Fock-state output statistics. For bunched output states (e.g.\ $|20000\rangle$), raw coincidence counts are multiplied by two to correct for the $50\%$ probability that the PNR MZI splits both photons across its output ports rather than directing them to the same channel. The HOM visibility of $0.991$ confirmed in the state-preparation unit (Fig.~\ref{fig:experimental_setup}, inset) verifies near-ideal photon indistinguishability under the operating conditions of all piston experiments.

\subsection*{Results overview and measurement protocols}
We perform four complementary measurement protocols spanning expansion and compression strokes across the full adiabatic-to-non-adiabatic crossover. All parameters follow the natural-unit convention of Ref.~\cite{Liu2024} ($\hbar = k_B = M = 1$), so that energies are in units of $\pi^2/2$ and temperature $T$ is in units of energy. In each protocol the system is prepared in a thermal Gibbs state at temperature $T$ and length $\lambda_0$, after which the bath is decoupled and the wall displaced under unitary evolution. The temperature is chosen so that at $\lambda_0$ both bosons occupy the lowest four levels with probability exceeding $95\%$, ensuring self-consistency with the four-level truncation. The four protocols are: (i)~\emph{velocity sweep, expansion}: $\lambda_0=1.0\to\lambda_\tau=3.0$, $v=0.1$--$6.0$, $T=5.0$; (ii)~\emph{velocity sweep, compression}: $\lambda_0=3.0\to\lambda_\tau=1.0$, same velocity range, $T=0.5$; (iii)~\emph{final-length sweep, expansion}: $v=1.1$, $\lambda_0=1.0$, $\lambda_\tau$ varied, $T=5.0$; (iv)~\emph{final-length sweep, compression}: $v=-0.7$, $\lambda_0=5.0$, $\lambda_\tau$ varied, $T=0.3$.

The residual unitary error
\begin{equation}
    \epsilon_{\rm unitary}(\%) = 100\,|1 - F|, \qquad
    F = \frac{1}{d}\left|\mathrm{Tr}\!\left(\sqrt{U_{5\times5}
    U_{5\times5}^\dagger}\right)\right|,
    \label{eq:unitary_error}
\end{equation}
grows with $|v|$ and $\lambda_\tau$ as leakage beyond $E_4$ increases. In the results presented here, we consider only protocols for which $\epsilon_{\rm unitary}<5\%$.

\begin{figure}[t]
  \centering
  \includegraphics[width=0.9\linewidth]{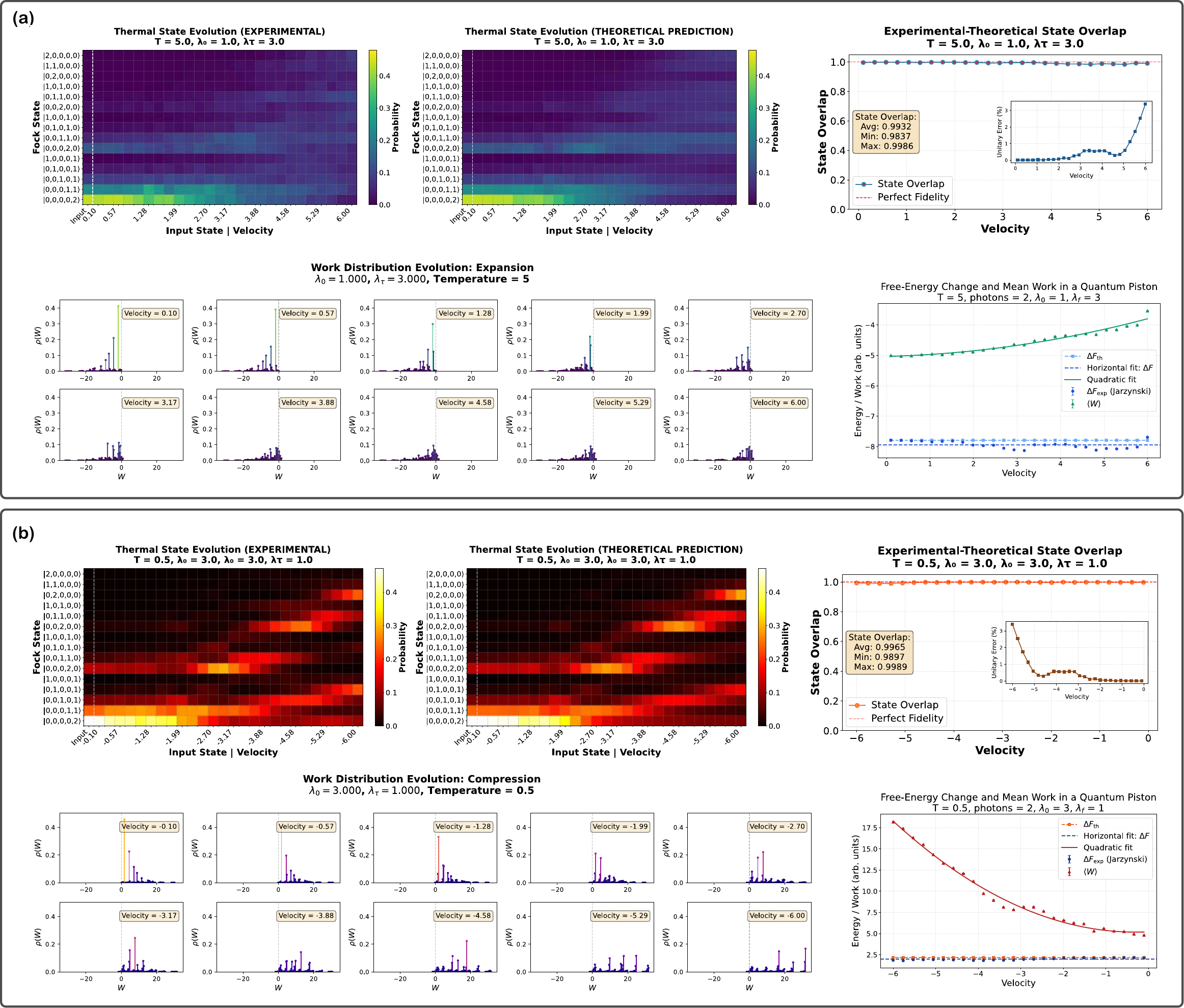}
\caption{\textbf{Thermal two-photon state evolution and thermodynamic signatures under
velocity-swept expansion and compression driving.} Panel (\textbf{a}) shows expansion from $\lambda_0=1.0$ to $\lambda_\tau=3.0$ at $T=5.0$, and panel (\textbf{b}) shows compression from $\lambda_0=3.0$ to $\lambda_\tau=1.0$ at $T=0.5$. In each panel, the top-left and top-middle maps present the experimentally reconstructed and theoretically predicted two-photon output probability distributions in the Fock basis as the piston velocity is swept. Fock states are ordered along the vertical axis in ascending order of total energy, with the lowest-energy state at the bottom. The top-right plot reports the state overlap (Bhattacharyya coefficient), with the dashed reference line indicating unity. The inset shows the unitary error of the implemented $5\times5$ quasi-unitary matrices, quantifying residual non-unitarity. The lower-left block displays velocity-resolved thermal work distributions. The lower-right plot summarizes free-energy/work metrics: mean work $\langle W\rangle=\sum_W p(W)\,W$, experimental free-energy estimate $\Delta F_{\mathrm{exp}}=-T\ln\langle e^{-W/T}\rangle$ (Jarzynski estimator), and theoretical free-energy change $\Delta F_{\mathrm{th}}=-T\ln\!\big[Z(\lambda_\tau)/Z(\lambda_0)\big]$ from equilibrium partition functions. In the velocity-sweep analysis, $\Delta F_{\mathrm{exp}}$ is additionally shown with a horizontal fit, while $\langle W\rangle$ is fitted by a quadratic function of $|v|$ (labeled \textit{Quadratic fit}), capturing the increasing work free-energy gap with driving speed. $\Delta F_{\mathrm{th}}$ is shown as a dashed connected trend. For each branch, the temperature is chosen so that at $\lambda_0$ the probability that both bosons occupy the first four energy levels exceeds $95\%$, consistent with the four-level truncation used in the analysis. All quantities are reported in natural units with $k_B=1$, boson mass $M=1$, and $\hbar=1$.}
  \label{fig:velocity_sweep}
\end{figure}

\subsection*{Velocity-dependent thermal dynamics and Jarzynski validation}

We map the evolution of the thermal two-photon Gibbs state under finite-time piston driving across the full adiabatic-to-non-adiabatic crossover. Fig.~\ref{fig:velocity_sweep}(a) shows the expansion protocol ($\lambda_0 = 1.0 \to \lambda_\tau = 3.0$, $k_BT = 5.0$) and Fig.~\ref{fig:velocity_sweep}(b) the compression protocol ($\lambda_0 = 3.0 \to \lambda_\tau = 1.0$, $k_BT = 0.5$). In each panel, the top-left and top-middle heatmaps display the experimentally reconstructed and theoretically predicted two-photon Fock-state output distributions $P(\mathbf{n}_f | v)$ as the wall velocity is swept~\cite{Liu2024}; the top-right plot reports the Bhattacharyya coefficient $B \in [0,1]$ quantifying experimental--theoretical agreement ($B=1$ indicates perfect overlap; see Methods for details)~\cite{Fuchs1999}, and the inset shows $\epsilon_{\rm unitary}$ for each implemented transformation. The lower-left panels display the velocity-resolved thermal work distributions $P(W|v)$, and the lower-right panels summarize the key thermodynamic metrics discussed below. Within the two-projective-measurement scheme, the thermal work distribution is given by
\begin{equation}
    P(W) = \sum_{\mathbf{n}_i, \mathbf{n}_f} 
    \frac{e^{-E_{\mathbf{n}_i}(\lambda_0)/T}}{Z_0}\,
    P(\mathbf{n}_f | \mathbf{n}_i)\,
    \delta\!\left(W - E_{\mathbf{n}_f}(\lambda_\tau) + 
    E_{\mathbf{n}_i}(\lambda_0)\right),
    \label{eq:work_dist}
\end{equation}
where $E_{\mathbf{n}}(\lambda) = \sum_k n_k\,k^2/\lambda^2$ is the total two-boson energy of Fock state $|\mathbf{n}\rangle$ at box length $\lambda$ (natural units $\hbar = k_B = M = 1$), $Z_0$ is the initial partition function, and $P(\mathbf{n}_f|\mathbf{n}_i)$ is the experimentally measured conditional output probability for input $|\mathbf{n}_i\rangle$. Output probabilities at work values differing by less than a numerical tolerance are grouped following the GPE method of Ref.~\cite{Liu2024}.

\paragraph{Expansion (Fig.~\ref{fig:velocity_sweep}(a)).}
At low velocities ($|v| \lesssim 1$), the driving is quasi-adiabatic: the quantum adiabatic theorem~\cite{Kato1950} guarantees that the system remains in its instantaneous eigenstate throughout the protocol, so the output distribution closely mirrors the initial thermal state and each boson stays in its occupied level as the box expands. The experimental and theoretical heatmaps are in excellent agreement throughout the full velocity range, with Bhattacharyya overlaps averaging $0.993$ and remaining above $0.984$ at all velocities (top-right panel of Fig.~\ref{fig:velocity_sweep}(a)), confirming that the implemented unitaries faithfully encode the piston dynamics across the entire adiabatic-to-non-adiabatic crossover. Correspondingly, $P(W|v)$ is concentrated near discrete negative work values $W = E_{\mathbf{n}_i}(\lambda_\tau) - E_{\mathbf{n}_i}(\lambda_0) < 0$, reflecting the energy decrease of each level under expansion ($E_k \propto 1/\lambda^2$), with near-zero probability of positive work (lower-left panel of Fig.~\ref{fig:velocity_sweep}(a)). As $|v|$ increases, non-adiabatic inter-level transitions become significant: bosons are promoted to higher levels $n_f > n_i$ whose energies at $\lambda_\tau$, despite the overall decrease in level spacing, can exceed the initial energy $E_{n_i}(\lambda_0)$, generating positive-work outcomes and broadening $P(W|v)$ into a multi-valued structure spanning both negative and positive work values. For two identical bosons, these transition amplitudes involve permanents of submatrices of the single-particle propagator, further enhancing the redistribution relative to distinguishable particles~\cite{Gong2014}.

The lower-right panel of Fig.~\ref{fig:velocity_sweep}(a) summarizes three thermodynamic quantities as a function of $|v|$. The mean work $\langle W \rangle = \sum_W P(W)\,W$ grows monotonically with $|v|$ and systematically exceeds the equilibrium free-energy difference
\begin{equation}
    \Delta F_{\rm th} = -T \ln\!\left[\frac{Z(\lambda_\tau)}{Z(\lambda_0)}\right],
    \label{eq:dFth}
\end{equation}
where $Z(\lambda) = \sum_{\mathbf{n}} e^{-E_{\mathbf{n}}(\lambda)/T}$ is the two-boson partition function evaluated over the four resolved levels. The inequality $\langle W \rangle \geq \Delta F_{\rm th}$, guaranteed by the second law of thermodynamics, holds at all driving speeds~\cite{Jarzynski1997}. The excess $W_{\rm diss} = \langle W \rangle - \Delta F_{\rm th} \geq 0$ quantifies the dissipated work, directly connected to irreversible entropy production $\Delta S_{\rm irr} = W_{\rm diss}/T \geq 0$, which vanishes only in the quasistatic limit. Since the bath is decoupled before the stroke and never reconnected, even at the lowest $|v|$ populations remain frozen at their $\lambda_0$ values while the equilibrium distribution shifts to $\lambda_\tau$, generating nonzero entropy production at all velocities. The gap $W_{\rm diss}$ grows further with $|v|$ as non-adiabatic excitations contribute additional irreversibility beyond this baseline.

The Jarzynski equality~\cite{Jarzynski1997} states that for any protocol initiated from thermal equilibrium,
\begin{equation}
    \left\langle e^{-W/T} \right\rangle = e^{-\Delta F_{\rm th}/T},
    \label{eq:jarzynski}
\end{equation}
regardless of driving speed or whether the final state is in equilibrium. The Jarzynski free-energy estimator
\begin{equation}
    \Delta F_{\rm exp} = -T\ln\!\left\langle e^{-W/T}\right\rangle
    \label{eq:jarzynski_estimator}
\end{equation}
recovers $\Delta F_{\rm th}$ for any driving speed. As shown in the lower-right panel, $\Delta F_{\rm exp}$ remains consistent with $\Delta F_{\rm th}$ to within $0.07T$ across the full velocity sweep, providing experimental confirmation of Jarzynski's equality within the four-level resolved subspace. While $\langle W \rangle$ grows monotonically with $|v|$ and is well described by a quadratic fit, $\Delta F_{\rm exp}$ stays pinned to $\Delta F_{\rm th}$ because the rare low-work trajectories exponentially reweighted by $e^{-W/T}$ precisely compensate for the average excess, a hallmark of the fluctuation theorem even under strongly irreversible driving~\cite{Crooks1999}.

\paragraph{Compression (Fig.~\ref{fig:velocity_sweep}(b)).}
At $T = 0.5$ and $\lambda_0 = 3.0$, the widely spaced energy levels combined with the low temperature mean that the initial Gibbs state is strongly concentrated in the lowest Fock components, with over $83\%$ of the thermal weight in the three lowest states $|02000\rangle$, $|01100\rangle$, and $|00200\rangle$. Compression to $\lambda_\tau = 1.0$ increases all energy levels by a factor of nine ($E_k \propto 1/\lambda^2$), so even in the quasi-adiabatic limit the work values are large and positive, with the mean adiabatic work ($\sim 5T$ above $\Delta F_{\rm th}$) reflecting the significant irreversibility inherent to this protocol even without inter-level transitions. As $|v|$ increases, faster compression enhances non-adiabatic inter-level coupling because the level spacing grows rapidly as the wall moves inward, populating higher Fock states and generating a strongly asymmetric, positively skewed $P(W|v)$ with a pronounced high-work tail (lower-left panel of Fig.~\ref{fig:velocity_sweep}(b)). For two identical bosons, the permanent-enhanced transition amplitudes further amplify this redistribution relative to distinguishable particles~\cite{Gong2014}. The Fock-state heatmaps show excellent experimental--theoretical agreement throughout, with Bhattacharyya overlaps averaging $0.997$ across the full velocity range. As shown in the lower-right panel, $\langle W \rangle$ grows monotonically with $|v|$ and is well described by a quadratic fit, while $\Delta F_{\rm exp}$ accurately recovers $\Delta F_{\rm th}$ across the full sweep, confirming the robustness of Jarzynski's equality Eq.~(\ref{eq:jarzynski}) even under the large
irreversibility of strongly driven compression strokes.

\begin{figure}[t]
  \centering
  \includegraphics[width=0.9\linewidth]{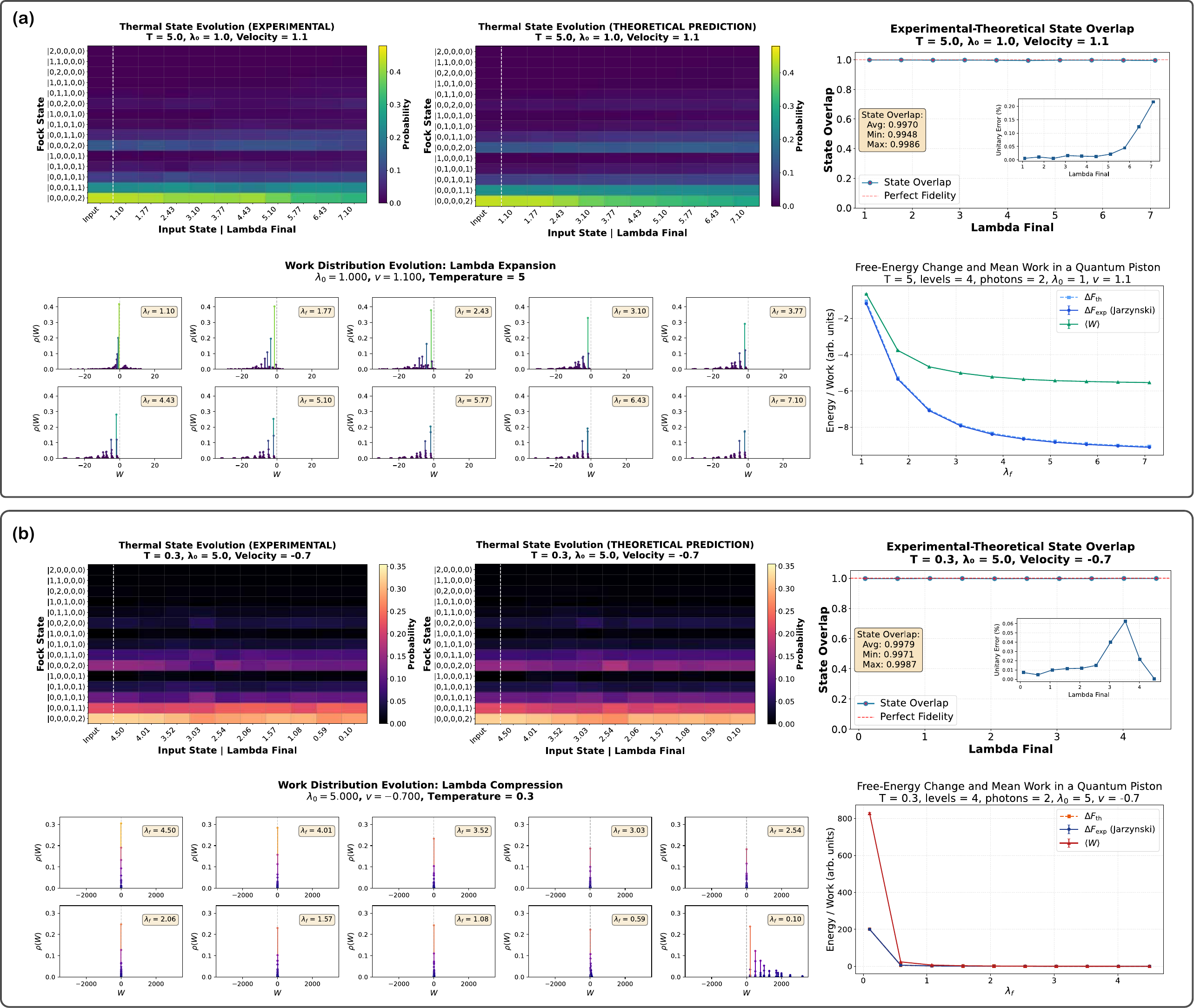}
  
\caption{\textbf{Thermal two-photon state evolution and thermodynamic metrics versus final
trap parameter.} Panel (\textbf{a}) shows the $\lambda$-sweep expansion branch with fixed velocity $v=1.1~\mathrm{mm/s}$, $\lambda_0=1.0$, and $T=5.0$, while panel (\textbf{b}) shows the $\lambda$-sweep compression branch with fixed velocity $v=-0.7~\mathrm{mm/s}$, $\lambda_0=5.0$, and $T=0.3$. In each panel, the top-left and top-middle heatmaps are the experimental and theoretical two-photon output probabilities in the Fock basis as $\lambda_\tau$ is varied. Fock states are ordered along the vertical axis in ascending order of total energy, with the lowest-energy state at the bottom. The top-right plot gives the experimental--theoretical state overlap (Bhattacharyya coefficient), with the dashed line indicating perfect fidelity. Insets report the unitary error of the implemented $5\times5$ quasi-unitary matrices, quantifying residual non-unitarity associated with representing leakage from the truncated four-level subspace with a single ancilla mode. The lower-left block shows the corresponding work distributions at representative $\lambda_\tau$ points, and the lower-right plot summarizes $\langle W\rangle=\sum_W p(W)\,W$, $\Delta F_{\mathrm{exp}}=-T\ln\!\langle e^{-W/T}\rangle$ (Jarzynski estimator), and $\Delta F_{\mathrm{th}}=-T\ln\!\big[Z(\lambda_\tau)/Z(\lambda_0)\big]$. Temperatures are chosen so that, at $\lambda_0$, the probability that both bosons occupy the first four energy levels exceeds $95\%$, ensuring consistency with the four-level truncation. All quantities are in natural units with $k_B=1$, boson mass $M=1$, and $\hbar=1$.}
\label{fig:lambda_sweep}
\end{figure}

\subsection*{Final trap parameter sweeps at fixed driving speed}
We next fix the driving speed and systematically vary $\lambda_\tau$ to isolate geometric effects from dynamical ones (Fig.~\ref{fig:lambda_sweep}).

\paragraph{Expansion (Fig.~\ref{fig:lambda_sweep}(a)).}
With $v = 1.1$, $\lambda_0 = 1.0$, and $T = 5.0$ fixed, $\lambda_\tau$ is increased from $1.1$ to $7.1$. The experimental and theoretical heatmaps are in excellent agreement throughout, with Bhattacharyya overlaps averaging $0.997$ and remaining above $0.995$ across all $\lambda_\tau$ values (top-right panel of Fig.~\ref{fig:lambda_sweep}(a)). The output distribution shifts gradually with $\lambda_\tau$: the probability of the lowest Fock state $|02000\rangle$ decreases progressively as $\lambda_\tau$ increases, reflecting growing non-adiabaticity at fixed $v$. The minimum energy gap $\Delta E \propto 1/\lambda_\tau^2$ shrinks with increasing $\lambda_\tau$, making inter-level transitions increasingly likely and redistributing population toward higher Fock states.

Inter-level transitions are already present at the smallest $\lambda_\tau = 1.1$. For an infinite square well with a moving wall, the dimensionless adiabaticity parameter for the $n\to m$ transition is~\cite{Doescher1969}
\begin{equation}
    \xi_{nm} = \frac{v\,|\langle m|\partial_\lambda n\rangle|}{E_m - E_n}
    = \frac{2mn\,v\,\lambda}{(m^2-n^2)^2},
    \label{eq:adiab}
\end{equation}
where $\langle m|\partial_\lambda n\rangle = 2mn/[\lambda(m^2-n^2)]$ quantifies how rapidly the eigenstates rotate in Hilbert space as the box expands, and $\xi_{nm} \ll 1$ is required for adiabatic evolution. For the lowest transition $n=1\to m=2$ at $v=1.1$, $\xi_{12} \approx 0.54$ at $\lambda_\tau = 1.1$ (intermediate regime) and grows to $3.47$ at $\lambda_\tau = 7.1$ (strongly non-adiabatic). Since the level energies at $\lambda_\tau = 1.1$ have decreased by only $\sim\!17\%$ relative to $\lambda_0$, a boson promoted to a moderately higher level can still carry more energy than it started with, yielding a small but nonzero $P(W>0)$. As $\lambda_\tau$ increases further, all levels compress as $E_k \propto 1/\lambda_\tau^2$: beyond $\lambda_\tau \approx 3.1$, even promotion to the highest accessible level cannot produce $W > 0$, and $P(W>0)$ vanishes entirely.

As $\lambda_\tau$ grows, both $\Delta F_{\rm th}$ and $\langle W \rangle$ decrease monotonically with a saturating rate as all energy levels become negligibly small. The dissipated work $W_{\rm diss} = \langle W \rangle - \Delta F_{\rm th}$ grows from $0.08T$ at $\lambda_\tau = 1.1$ to $0.70T$ at $\lambda_\tau = 7.1$, driven primarily by the increasing equilibrium entropy change: even in the adiabatic limit, populations frozen at $\lambda_0$ fall progressively further from the equilibrium distribution at $\lambda_\tau$. Despite this growth, $\Delta F_{\rm exp}$ tracks $\Delta F_{\rm th}$ to within $0.03T$ across the full sweep, confirming Jarzynski's equality as a function of geometry at fixed driving speed.

\paragraph{Compression (Fig.~\ref{fig:lambda_sweep}(b)).}
With $v = -0.7$, $\lambda_0 = 5.0$, and $T = 0.3$ fixed, $\lambda_\tau$ is decreased from $4.5$ to $0.1$. The heatmaps show excellent agreement throughout, with Bhattacharyya overlaps averaging $0.998$ and remaining above $0.997$ (top-right panel of Fig.~\ref{fig:lambda_sweep}(b)). In contrast to the expansion case, the probability of the lowest output Fock state $|02000\rangle$ increases gradually as $\lambda_\tau$ decreases, reflecting growing adiabaticity: $\xi_{12}$ decreases from $1.40$ at $\lambda_\tau = 4.5$ to $0.03$ at $\lambda_\tau = 0.1$ as the energy gap $\Delta E \propto 1/\lambda_\tau^2$ grows rapidly under compression, increasingly suppressing inter-level transitions.

Despite this growing adiabaticity, both $\langle W \rangle$ and $\Delta F_{\rm th}$ exhibit a dramatic increase at small $\lambda_\tau$, rising to $2757T$ and $668T$ respectively at $\lambda_\tau = 0.1$. This is driven entirely by the $1/\lambda_\tau^2$ divergence of the energy levels: the work associated with each output Fock state scales as $W \sim E_k(\lambda_\tau) \propto 1/\lambda_\tau^2$, so even a modest decrease in $\lambda_\tau$ near zero produces an enormous increase in work values. For instance, the adiabatic work for the dominant state $|02000\rangle$, the work done when both bosons remain in their initial levels, grows from $0.1T$ at $\lambda_\tau = 4.5$ to $666T$ at $\lambda_\tau = 0.1$, with a $35\times$ jump between $\lambda_\tau = 0.59$ and $\lambda_\tau = 0.1$ alone. The dissipated work $W_{\rm diss}$ similarly grows from $0.04T$ at $\lambda_\tau = 4.5$ to $59.6T$ at $\lambda_\tau = 0.59$, driven by the same energy scaling.. Despite these work distributions spanning orders of magnitude, $\Delta F_{\rm exp}$ recovers $\Delta F_{\rm th}$ to within $0.20T$ across the full sweep, providing a stringent confirmation of Jarzynski's equality under extreme compression conditions.

\begin{figure}[t]
  \centering
  \includegraphics[width=\linewidth]{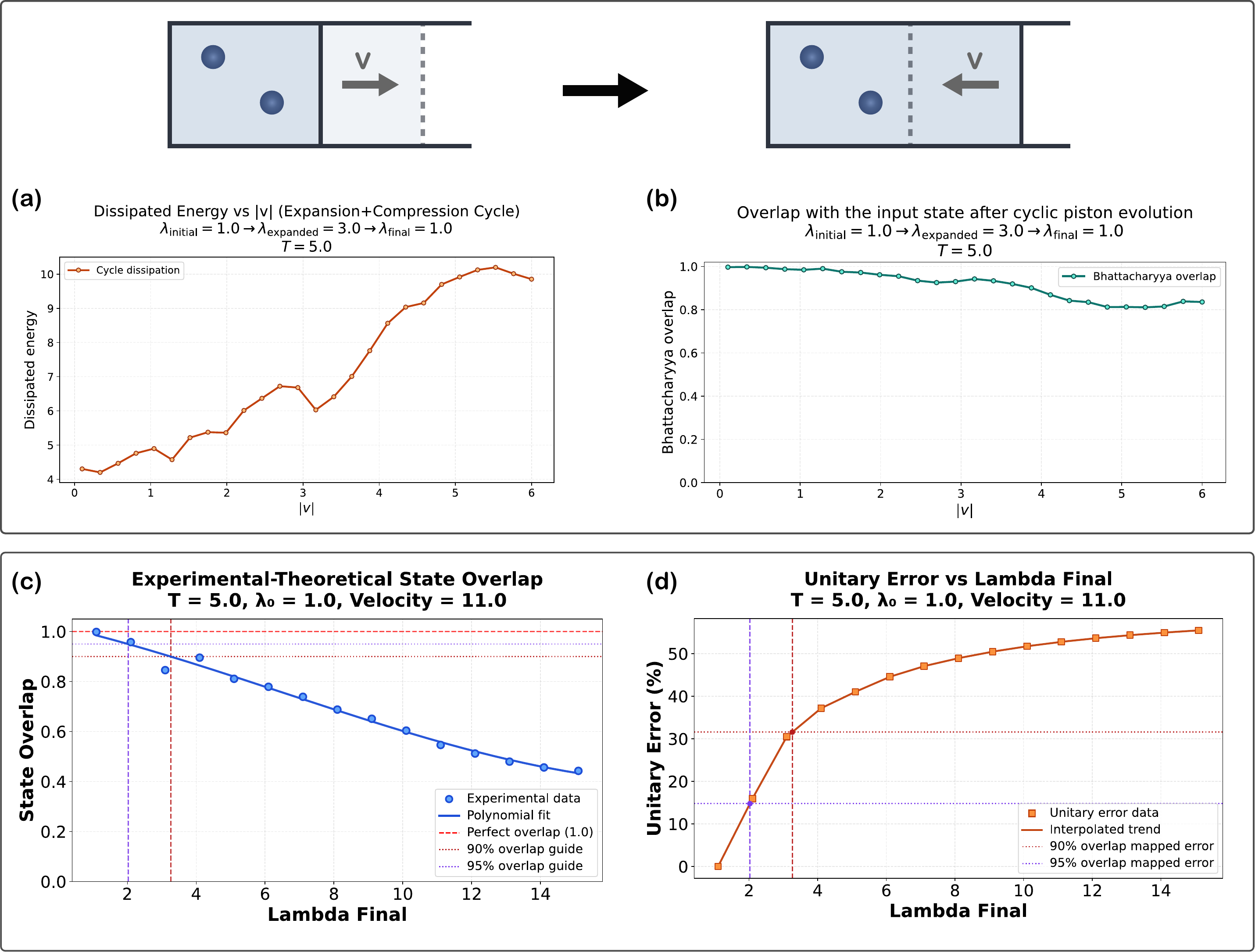}
\caption{\textbf{Cyclic quantum-piston protocol and error propagation analysis.} \textbf{(a)} Schematic of one expansion--compression cycle, from $\lambda_{\mathrm{initial}}=1.0$ to $\lambda_{\mathrm{expanded}}=3.0$ and back to $\lambda_{\mathrm{final}}=1.0$, at T = 5.0, with the corresponding cycle-dissipated energy versus $|v|$. \textbf{(b)} Bhattacharyya overlap with the input state after the full cycle versus $|v|$. \textbf{(c)} Experimental--theoretical state overlap versus $\lambda_{\mathrm{final}}$ (velocity $=11.0$), with a polynomial fit and 90\%/95\% overlap guide levels used to identify threshold operating points. \textbf{(d)} Unitary error versus $\lambda_{\mathrm{final}}$, where the thresholds from panel (c) are mapped onto this curve to quantify how unitary-level imperfections translate into overlap degradation; i.e., panel (d) provides a direct mapping from unitary error to the overlap error it can induce.}
\label{fig:cyclic_protocol_error_mapping}
\end{figure}

\subsection*{Full-cycle irreversibility and dissipation vs.\ geometry}
To probe closed thermodynamic cycles, we chain an expansion stroke ($\lambda_0 = 1.0 \to \lambda_\tau = 3.0$) with its time-reversed compression back to $\lambda_0 = 1.0$, using the evolved output state of the expansion as the input to the compression stroke without any intermediate thermal resetting [Fig.~\ref{fig:cyclic_protocol_error_mapping}(a)]. The total dissipated work per cycle,
\begin{equation}
W_{\mathrm{diss,cyc}} = \bigl(\langle W\rangle_{\rm exp} - \Delta F_{\rm exp}\bigr)
+
\bigl(\langle W\rangle_{\rm comp} - \Delta F_{\rm comp}\bigr),
\end{equation}
is nonzero even at the lowest driving speed: the thermal ensemble always produces a work distribution with finite variance, so by Jensen's inequality $W_{\rm diss,cyc} > 0$ unless $P(W)$ is a delta function~\cite{Jarzynski1997}. Beyond this baseline, $W_{\rm diss,cyc}$ grows monotonically with $|v|$ as non-adiabatic excitations broaden $P(W)$ and contribute additional irreversibility.

The Bhattacharyya overlap between the initial thermal state and the state after one complete cycle,
\begin{equation}
B_{\rm cycle} = \sum_{\mathbf{n}}
\sqrt{p_{\rm initial}(\mathbf{n})\,p_{\rm final}(\mathbf{n})},
\end{equation}
provides a complementary, dynamical measure of irreversibility [Fig.~\ref{fig:cyclic_protocol_error_mapping}(b)]: while $W_{\rm diss,cyc}$ captures the statistical spread of work values, $B_{\rm cycle}$ measures whether the Fock-state populations return to their initial values. At low $|v|$, quasi-adiabatic evolution returns the distribution nearly intact ($B_{\rm cycle} > 0.99$ for $|v| \lesssim 0.5$), consistent with a near-zero dynamical contribution to irreversibility. At higher speeds, $B_{\rm cycle}$ degrades systematically, dropping below $0.95$ at $|v| \approx 2.5$ and below $0.90$ at $|v| \approx 4.1$, as non-adiabatic transitions populate higher Fock components during expansion that cannot fully de-excite during compression~\cite{Gong2014}.

To establish the operational limits of the photonic implementation, we map the experimental--theoretical Bhattacharyya overlap onto $\epsilon_{\rm unitary}$ by sweeping $\lambda_\tau$ at fixed $|v| = 11.0$, exploiting the monotonic growth of embedding error with stroke length. As shown in Fig.~\ref{fig:cyclic_protocol_error_mapping}(c,d), a $95\%$ state overlap requires $\epsilon_{\rm unitary} \lesssim 14.8\%$, while a $90\%$ overlap tolerates up to $\epsilon_{\rm unitary} \approx 31.6\%$, providing concrete hardware benchmarks for the fidelity achievable at a given embedding error.

\section*{Conclusion}
We have experimentally demonstrated the simulation of nonequilibrium quantum thermodynamics of a two-boson quantum piston on a programmable integrated photonic processor. By encoding the piston propagator via a quasi-unitary embedding in a $12\times12$ Clements interferometer and reconstructing full two-photon work distributions, we have shown that bosonic indistinguishability encoded in the permanent structure of the many-body transition amplitudes fundamentally reshapes quantum work statistics relative to distinguishable particles, observed the full adiabatic-to-non-adiabatic crossover for both expansion and compression protocols, and experimentally verified the Jarzynski equality across a wide range of driving speeds and trap geometries. For cyclic protocols, we have further quantified irreversibility through dissipated work and state-return fidelity, and established concrete hardware benchmarks linking embedding error to implementation fidelity.

The significance of our platform extends beyond this two-boson proof of principle. The many-boson quantum piston work distribution is determined by permanents of single-particle transition-amplitude matrices~\cite{Gong2014}, a quantity that is $\#P$-hard to compute classically~\cite{Aaronson2011}. Photonic processors therefore offer a natural and scalable route to accessing thermodynamic work statistics in regimes that are provably intractable for classical simulation as the boson number grows. Our architecture is fully programmable, enabling systematic study of how bosonic interference, particle number, driving protocol geometry, and temperature together determine quantum dissipation and entropy production - questions of direct relevance to the design of quantum heat engines~\cite{Gong2014}, refrigerators, and thermodynamic machines operating in the genuinely quantum many-body regime~\cite{Jaramillo2016, Myers2020}.

\section*{Methods}

\subsection*{Single-particle piston transition amplitudes and numerical implementation}

Although the experiment involves multiple indistinguishable bosons, the piston dynamics can be fully characterized by a single-particle transformation matrix. This is because the bosons are non-interacting, so the many-body Hamiltonian is a sum of identical single-particle Hamiltonians and the total unitary evolution factorizes accordingly. In this case, the evolution of creation operators is linear, and the many-body transformation in the Fock basis is completely determined by the single-particle unitary matrix. Multi-boson transition amplitudes are obtained by bosonic symmetrization, i.e., as matrix permanents of submatrices of the single-particle transformation. Therefore, the single-particle piston matrix serves directly as the linear transformation implemented in the boson-sampling framework.

We consider a particle of mass $M$ confined in a one-dimensional infinite square well of time-dependent length $\lambda(t)$ with hard-wall boundary conditions at $x=0$ and $x=\lambda(t)$. The Hamiltonian is
\begin{equation}
\hat H(t) = -\frac{\hbar^2}{2m}\frac{\partial^2}{\partial x^2},
\qquad 0 < x < \lambda(t),
\end{equation}
where $\hbar$ is the reduced Planck constant. For a fixed length $\lambda$, the normalized eigenfunctions are
\begin{equation}
\phi_n(x;\lambda)
=
\sqrt{\frac{2}{\lambda}}
\sin\!\left(\frac{n\pi x}{\lambda}\right),
\end{equation}
with corresponding eigenenergies
\begin{equation}
E_n(\lambda)
=
\frac{\hbar^2\pi^2 n^2}{2m\lambda^2},
\end{equation}
where $n=1,2,\dots$ labels the energy level and $x\in(0,\lambda)$.

We consider a linear expansion/compression protocol
\begin{equation}
\lambda(t)=\lambda_0+vt,
\end{equation}
where $\lambda_0$ and $\lambda_\tau$ are the initial and final lengths, $v$ is the wall velocity, and the protocol duration is
\begin{equation}
\tau=\frac{\lambda_\tau-\lambda_0}{v}.
\end{equation}

The time-dependent Schr\"odinger equation
\begin{equation}
i\hbar \frac{\partial}{\partial t}\Psi(x,t)=\hat H(t)\Psi(x,t)
\end{equation}
admits an exact orthogonal solution set for linear wall motion,
\begin{equation}
\Phi_j(x,t)
=
\exp\!\left[
\frac{i}{\hbar\,\lambda(t)}
\left(
\frac{1}{2} m v x^2
-
E_j(\lambda_0)\lambda_0 t
\right)
\right]
\phi_j(x;\lambda(t)),
\end{equation}
where $j=1,2,\dots$ labels the mode. If the system is initially prepared in eigenstate $\Psi(x,0)=\phi_i(x;\lambda_0)$, the time-evolved wavefunction can be expanded as
\begin{equation}
\Psi(x,t)=\sum_{j=1}^{\infty} c_{ji}\,\Phi_j(x,t),
\end{equation}
with coefficients determined by the initial overlap,
\begin{equation}
c_{ji}
=
\frac{2}{\lambda_0}
\int_0^{\lambda_0}
\exp\!\left(
-\frac{i m v x^2}{2\hbar\lambda_0}
\right)
\sin\!\left(\frac{j\pi x}{\lambda_0}\right)
\sin\!\left(\frac{i\pi x}{\lambda_0}\right)
dx .
\end{equation}

The transition amplitude from initial level $i$ to final level $f$ is
\begin{equation}
T_{fi}
=
\langle f^{\lambda_\tau}|\hat U(\tau)|i^{\lambda_0}\rangle
=
\int_0^{\lambda_\tau}
\phi_f^*(x;\lambda_\tau)\Psi(x,\tau)\,dx
=
\sum_{j=1}^{\infty}
c_{ji}
\int_0^{\lambda_\tau}
\Phi_j(x,\tau)\phi_f^*(x;\lambda_\tau)\,dx,
\end{equation}
where $\hat U(\tau)$ is the time-evolution operator and $f$ labels the final eigenstate~\cite{Liu2024}.

In practice, the infinite summation over $j$ is truncated at $j\le j_{\max}$, chosen such that $\sum_{j>j_{\max}}|c_{ji}|^2\ll1$. The truncated single-particle matrix $T_{N\times N}$ is obtained by restricting $i,f\in\{1,\dots,N\}$, with $N=4$ in the experiment.

\paragraph{Numerical implementation.}

The transition amplitudes $T_{fi}$ are evaluated numerically from the analytical expressions above. The spatial integrals defining the overlap coefficients and final projections are computed using adaptive numerical integration with controlled convergence. The infinite mode summation is truncated at $j_{\max}=50$, which ensures that increasing the cutoff further does not change the transition amplitudes within numerical precision. Unless otherwise specified, calculations are performed in natural units with $m=\hbar=k_B=1$. The truncated single-particle transformation matrix $T_{N\times N}$ is obtained by evaluating $T_{fi}$ for $i,f\in\{1,\dots,N\}$, with $N=4$ in the experiment.

\subsection*{Quasiunitary extension of the truncated piston transformation}

The truncated single-particle piston transformation $T_{4\times4}$ is generally non-unitary because population can leak into higher piston energy levels outside the retained four-level manifold. Since passive linear optical interferometers implement unitary transformations, $T_{4\times4}$ must be embedded into a larger unitary operation acting on additional modes~\cite{Tischler2018, krishna2025emulationcoherentabsorptionquantum}.

We first perform the singular value decomposition (SVD)
\begin{equation}
T_{4\times4}=UDW,
\end{equation}
where $U$ and $W$ are $4\times4$ unitary matrices and
$D=\mathrm{diag}(d_1,d_2,d_3,d_4)$ contains singular values $0\le d_k\le1$. 
Values $d_k<1$ quantify population leakage from the truncated subspace along the corresponding singular directions.

In general, an exact unitary dilation requires
\begin{equation}
m_{\mathrm{anc}}=\mathrm{rank}\!\left(I-T^\dagger T\right)
\end{equation}
ancilla modes, equal to the number of singular values strictly smaller than unity.

In our implementation we restrict the extension to a single ancilla mode in order to maintain a compact realization within the available $12\times12$ interferometer. This ancilla represents the collective subspace of higher piston energy levels $\{E_5,E_6,\dots\}$ that lie outside the truncated basis. The enlarged transformation is therefore written as
\begin{equation}
U_{5\times5}=
\begin{pmatrix}
T_{4\times4} & \mathbf{b} \\
\mathbf{c}^\dagger & \alpha
\end{pmatrix},
\end{equation}
where the upper-left block acts on the four retained piston levels, the final row and column correspond to the ancilla mode, $\mathbf{b}$ and $\mathbf{c}$ describe coupling between the physical modes and the ancilla, and $\alpha$ acts within the ancilla channel.

Because only a single ancilla channel is available, exact unitarity cannot generally be achieved when
\begin{equation}
\mathrm{rank}\!\left(I-T_{4\times4}^\dagger T_{4\times4}\right)>1.
\end{equation}

The deviation from perfect unitarity is quantified using the fidelity-style metric employed in our numerical pipeline. Defining $\sigma = U_{5\times5}U_{5\times5}^\dagger$ and $d=5$, we compute
\begin{equation}
F=\frac{1}{d}\left|\mathrm{Tr}\!\left(\sqrt{\sigma}\right)\right|,
\end{equation}
and report the percentage unitary error as
\begin{equation}
\epsilon_{\mathrm{unitary}}(\%) = 100\,|1-F|.
\end{equation}

Only transformations exhibiting $\epsilon_{\mathrm{unitary}}<0.5\%$ are implemented experimentally.

\subsection*{Clements decomposition scheme}

The enlarged unitary matrix $U_{5\times5}$ is physically realized by decomposing it into a sequence of two-mode unitary operations using the rectangular Clements architecture~\cite{Clements2016}. In this framework, an arbitrary $N\times N$ unitary matrix is factorized into a mesh of nearest-neighbour SU(2) transformations arranged in alternating layers as shown in Fig.\ref{fig:chip_architecture}(a).

Each two-mode operation is implemented on-chip by a Mach--Zehnder interferometer (MZI), consisting of two fixed 50:50 multi-mode interference (MMI) beam splitters and two thermo-optic phase shifters: an internal phase $\theta$ and an external phase $\phi$. The transfer matrix of a single MZI is given by
\begin{equation}
U_{\mathrm{MZI}}(\theta,\phi)
=
e^{i\left(\frac{\theta}{2}+\frac{\pi}{2}\right)}
\begin{pmatrix}
e^{i\phi}\sin\!\left(\frac{\theta}{2}\right) & \cos\!\left(\frac{\theta}{2}\right) \\
e^{i\phi}\cos\!\left(\frac{\theta}{2}\right) & -\sin\!\left(\frac{\theta}{2}\right)
\end{pmatrix}.
\end{equation}

By arranging such MZIs in a rectangular mesh acting on adjacent spatial modes, the full target unitary transformation is synthesized.

For each piston parameter set $(v,\lambda_0,\lambda_\tau)$, the enlarged matrix 
$U_{5\times5}$ obtained from the quasiunitary extension is decomposed numerically 
into a set of MZI phase parameters $\{(\theta_k,\phi_k)\}$ corresponding to the 
ordered elements of the mesh. All piston transformations were programmed on 
\textbf{Noor-Q} by configuring the corresponding interferometer phases according 
to this decomposition.
The decomposition was implemented using a self-modified version of the open-source \texttt{interferometer} Python package (v1.1.1), installed via \texttt{pip} and adapted to match the transfer-matrix convention of our device. The modification ensures consistency between the analytical MZI model used in the decomposition and the physical phase convention of the fabricated chip.

In the present implementation, the active piston submesh requires ten MZIs arranged in four rectangular layers, sufficient to synthesize the full $5\times5$ unitary block within the selected mode subspace. All other MZIs of the $12\times12$ chip are configured in passive routing states ($\theta=\pi$ bar state or $\theta=0$ cross state), so that identity propagation is preserved outside the programmed region.

\subsection*{The photonic chip and thermal stabilization}
The programmable photonic circuit is fabricated at Advanced Micro Foundry on a silicon-on-insulator (SOI) platform, consisting of a 220\,nm thick silicon device layer above a 3\,$\mu$m buried oxide substrate. The 500\,nm wide waveguides support single-mode operation of the fundamental TE mode across the C-band (1530--1565\,nm), with typical propagation losses of 2\,dB/cm. Edge couplers at the chip facets provide fiber-to-chip coupling with losses of approximately 1.3\,dB for TE polarization. The chip integrates 264 elements (phase shifters and multimode interferometers) arranged as 12 input and 12 output waveguides interconnected through a reconfigurable mesh of thermally-tuned MZIs. Each MZI consists of two fixed 50:50 MMI beam splitters and two thermo-optic phase shifters providing independent control of the internal ($\theta$) and external ($\phi$) phase shifts. Deep trenches ($280\times12\,\mu$m) are etched around the MZIs to suppress thermal crosstalk between adjacent heaters.

The chip is mounted on a Peltier-cooled stage to suppress thermal drift and is packaged to ensure stable optical coupling to fiber arrays. Electrical currents driving the thermo-optic phase shifters are supplied via wire bonds connected to a custom printed circuit board (PCB). Each phase shifter is driven by a Qontrol Q8iv precision current source (740\,nA resolution), with active temperature stabilization provided by a Thorlabs TED200C PID controller using thermistor feedback, targeting a resistance of 10\,k$\Omega$ with stability within $\pm0.01\,^\circ$C. The SPDC crystal is also mounted on a similar temperature-controlled stage, allowing precise tuning of the phase-matching condition to generate wavelength-degenerate photon pairs. This stabilization ensures high photon indistinguishability, which is essential for high-fidelity two-photon interference.

Before each measurement run, the interferometer is calibrated using classical laser light to determine the current-phase mapping of all relevant MZIs~\cite{Alexiev2021, Lin2024}. Each MZI is characterized through two sequential steps: first, a resistance measurement sweeping the applied current to extract the heater's thermal coefficients; second, an optical power sweep yielding a sinusoidal modulation from which the thermo-optic efficiency and static phase offset are extracted. Together these determine the current required to realize any target phase, ensuring accurate programming of the enlarged unitary transformation implementing the piston dynamics. Additional details regarding the experimental setup can be found in the supplementary.

\subsection*{Detection and data acquisition}

Single-photon detection is performed using superconducting nanowire single-photon detectors (SNSPDs) manufactured by Single Quantum B.V., operated in a closed-cycle cryostat at 2.6\,K. All detectors exhibit system detection efficiencies exceeding 70\%, dark count rates below 10\,Hz, and timing jitter below 20\,ps. Detector dead times are negligible on the timescales relevant to the present coincidence measurements. Photon arrival times are recorded using a multi-channel Swabian Time-Tagger, enabling full reconstruction of coincidence events via post-processing. 

To account for unequal detection efficiencies across channels, a calibration procedure is performed by sequentially routing light to each output channel while maintaining all other conditions unchanged. The measured single-channel count rates are used to extract relative efficiency correction factors, which are applied to normalize the measured coincidence probabilities. This procedure compensates for detector-efficiency variations, fiber-coupling losses, and passive component imbalance.

Statistical uncertainties are calculated assuming Poissonian counting statistics, with error bars corresponding to $\sqrt{N}$ and propagated through the normalization procedure.

\subsection*{Multi-configuration detection strategy}

The photon-number-resolving (PNR) stage produces ten effective detection channels $\{d_1,\dots,d_{10}\}$ corresponding to the split outputs of the five piston modes. Two-photon statistics are determined by the set of pairwise coincidence observables

\begin{equation}
C_{ij} = \langle n_i n_j \rangle, \quad i<j,
\end{equation}

which in principle requires access to all $\binom{10}{2}$ detector pairs.

As only eight SNSPDs were available simultaneously, measurements were performed in three independent configurations $S_k$ ($k=1,2,3$), each containing eight detection channels ($|S_k|=8$). The configurations were chosen such that

\begin{equation}
\bigcup_{k=1}^{3} \{(i,j)\,|\, i,j \in S_k,\; i<j\}
=
\{(i,j)\,|\, i,j \in \{1,\dots,10\},\; i<j\},
\end{equation}

ensuring complete coverage of all required coincidence pairs.

For a fixed piston setting, coincidence counts $N^{(k)}_{ij}$ are recorded sequentially for each configuration while keeping the programmed interferometer unchanged. After correcting for relative channel efficiencies, the normalized
coincidence probabilities are

\begin{equation}
P_{ij} =
\frac{N_{ij}}{\sum_{i<j} N_{ij}},
\end{equation}

where $N_{ij}$ denotes the efficiency-corrected counts assembled from the appropriate configuration. The full set $\{P_{ij}\}$ is then mapped to the corresponding two-photon Fock-basis probabilities of the five piston modes.

\subsection*{The SPDC source}

Photon pairs are generated via type-II spontaneous parametric down-conversion (SPDC) in a periodically poled KTP (PPKTP) crystal pumped by a continuous-wave diode laser (Cobolt). The pump operates at a central wavelength of 785\,nm with an output power
of 70\,mW at the crystal input (drive current 220\,mA).

The down-converted photons are centered near 1570\,nm and are collected into single-mode fibers and routed to the integrated chip. When characterized directly at the source output prior to injection into the photonic circuit, single-photon detection rates of approximately 120\,kHz per channel are observed. The corresponding two-photon coincidence rate between signal and idler channels is approximately 20\,kHz at zero relative delay using a 2\,ns coincidence window, corresponding to a heralding efficiency of about 20\%. The single-photon character of the source was verified via a heralded second-order correlation measurement yielding $g^{(2)}(0) = 0.014$, with a maximum $g^{(2)}(\tau) = 0.027$ within the 2\,ns coincidence window; full spectral characterization of the photon pair, including joint spectral intensity measurements confirming the pair bandwidth and frequency anti-correlations, are reported in Ref.~\cite{krishna2025emulationcoherentabsorptionquantum}.

\subsection*{Theoretical calculation of output statistics} The programmable interferometer implements the enlarged single-photon unitary $U_{5\times5}$ obtained from the quasiunitary embedding of the truncated piston transition matrix. The matrix element $U_{ji}$ gives the probability amplitude for a photon entering mode $i$ to exit in mode $j$, so the single-photon output
probability is simply $|U_{ji}|^2$.

For two indistinguishable photons, the output statistics are governed by the bosonic scattering formalism: because photons are identical bosons, the two-photon transition amplitudes between input and output Fock states are given by permanents of $2\times2$ submatrices of $U_{5\times5}$, constructed by selecting rows and columns corresponding to the occupied input and output modes~\cite{Tichy2014}. The permanent is the bosonic analogue of the determinant, and its appearance here is the mathematical signature of many-body quantum interference between indistinguishable photons~\cite{Aaronson2011}. The two-photon output probability for each Fock state is the squared modulus of the corresponding permanent amplitude, normalized by the factorial of the mode occupation numbers.

For the five piston modes, the two-photon output space spans 15 Fock states. The mapping from $U_{5\times5}$ to the full two-photon transformation is computed using the open-source \texttt{qoptcraft} Python package (v2.0.0)~\cite{QOptCraft2022}. Theoretical distributions $\{q_i\}$ computed this way are directly compared with the experimentally reconstructed distributions $\{p_i\}$ using the Bhattacharyya coefficient defined in Eq.~(\ref{eq:bhattacharyya}).

\subsection*{Bhattacharyya coefficient}
The experimental--theoretical state overlap is quantified by the Bhattacharyya coefficient~\cite{Fuchs1999}
\begin{equation}
    B = \sum_{\mathbf{n}} \sqrt{P_{\rm expt}(\mathbf{n})\, P_{\rm th}(\mathbf{n})},
    \label{eq:bhattacharyya}
\end{equation}
where the sum runs over all two-photon Fock states $|\mathbf{n}\rangle$ in the five-mode output basis, $P_{\rm expt}(\mathbf{n})$ is the experimentally reconstructed output probability, and $P_{\rm th}(\mathbf{n})$ is the theoretically predicted probability. $B \in [0,1]$, with $B = 1$ indicating perfect agreement and $B = 0$ indicating orthogonal distributions.

\bmhead{Supplementary information}

Supplementary information is available for this paper, including additional details on the experimental setup, and additional experimental data.

\bmhead{Acknowledgements}

The authors thank Prof.\ Ivan M.\ Khaymovich, Prof.\ Zhang-Qi Yin and Prof.\ Xian-Min Jin for helpful discussions. The authors also thank Swabian Instruments for providing the multi-channel time-tagger used in this work, and Advanced Micro Foundry (AMF) for fabricating the photonic chip. We thank Assistant Professor Iman Esmaeil Zadeh for helpful discussions on superconducting nanowire single-photon detectors (SNSPDs). G.K.\ and A.W.E.\ acknowledge support from the Knut and Alice Wallenberg (KAW) Foundation through the Wallenberg Centre for Quantum Technology (WACQT). J.G.\ acknowledges support from the Swedish Research Council (Ref: 2023-06671 and 2023-05288), Vinnova (Ref: 2024-00466), and the G\"{o}ran Gustafsson Foundation. A.W.E.\ acknowledges support from the Swedish Research Council (VR) Starting Grant (Ref: 2016-03905) and the Vinnova quantum kick-start project 2021. V.Z.\ acknowledges support from the KAW Foundation.

\section*{Declarations}
The authors declare no competing financial interests.









\bibliography{sn-bibliography}

\end{document}


	\baselineskip18pt
	
	\section*{Supplementary Materials: Experimental simulation of non-equilibrium quantum piston on a programmable photonic quantum computer}

\noindent
\normalsize{Govind Krishna,$^{1,\ast}$ Rohan Yadgirkar,$^{1}$ Balakrishnan Krishnakumar,$^{1}$ Andrea Cataldo,$^{1}$\\ Ze-Sheng Xu,$^{1}$  Johannes W. N. Los,$^{2}$ Val Zwiller,$^{3}$\\ Jun Gao,$^{1,4\ast}$ \& Ali W. Elshaari$^{1,\ast}$}
\\
\\
\normalsize{$^1$Department of Applied Physics, KTH Royal Institute of Technology, Albanova University Centre, Roslagstullsbacken 21, 106 91 Stockholm, Sweden}\\
\normalsize{$^2$Single Quantum B.V., Delft 2628 CJ, The Netherlands}\\
\normalsize{$^3$RISE Research Institutes of Sweden, Stockholm, Sweden}\\
\normalsize{$^4$School of Optical and Electronic Information, Huazhong University of Science and Technology, Luoyu Road 1037, Wuhan, Hubei, 430074, China}\\

\normalsize{$^\ast$E-mail: govindk@kth.se, jungao@hust.edu.cn, elshaari@kth.se}\\

\baselineskip24pt

\section*{S1. Additional Experimental Setup Details}
\addtocounter{section}{1}

To provide further clarity on device implementation and control electronics, we include
the following elaborations.

\subsection*{S1.1 Fabrication and Layout}

\begin{figure}[t]
    \centering
    \includegraphics[width=\linewidth]{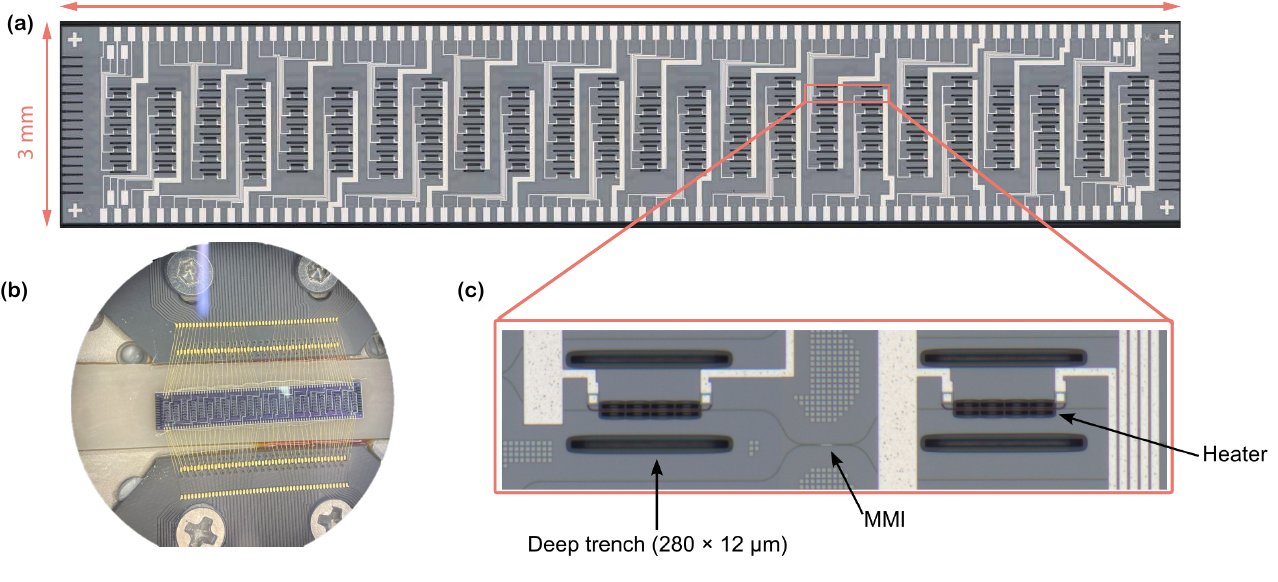}
    \caption{\textbf{Photonic chip characterization.}
    (\textbf{a}) Optical microscope image of the programmable photonic processor.
    The chip measures $16 \times 3$\,mm and integrates 264 elements (phase
    shifters and MMIs). (\textbf{b}) Photograph of the packaged chip mounted
    on the printed circuit board used for electrical interfacing. (\textbf{c})
    Magnified view of the MZI region, highlighting a multimode interference
    (MMI) coupler, a thermo-optic heater, and the surrounding deep trenches
    ($280 \times 12\,\mu$m) etched to reduce thermal crosstalk between
    neighbouring heaters.}
    \label{fig:chip_photo}
\end{figure}

Fig.~\ref{fig:chip_photo} provides optical micrographs of the fabricated chip, including a full-chip overview and a magnified view of the MZI region highlighting the deep trenches and thermo-optic heaters. The chip was fabricated at Advanced Micro Foundry (AMF) using their standard silicon-on-insulator (SOI) platform, with a 220\,nm thick silicon device layer above a 3\,$\mu$m buried oxide substrate. The 500\,nm wide waveguides support single-mode operation of the TE mode across the C-band (1530--1565\,nm). The thermo-optic phase shifters employ titanium nitride (TiN) resistive heaters patterned above the waveguide layer. Deep trenches ($280 \times 12\,\mu$m) are etched around each MZI to suppress thermal crosstalk between adjacent heaters. Edge couplers at the chip facets provide fiber-to-chip coupling with losses of approximately 1.3\,dB for TE polarization, and typical propagation losses are 2\,dB/cm. The full $12 \times 12$ Clements interferometer mesh integrates 264 elements (phase shifters and MMIs), providing 12 input and 12 output
waveguide ports.

\begin{figure}[t]
    \centering
    \includegraphics[width=\linewidth]{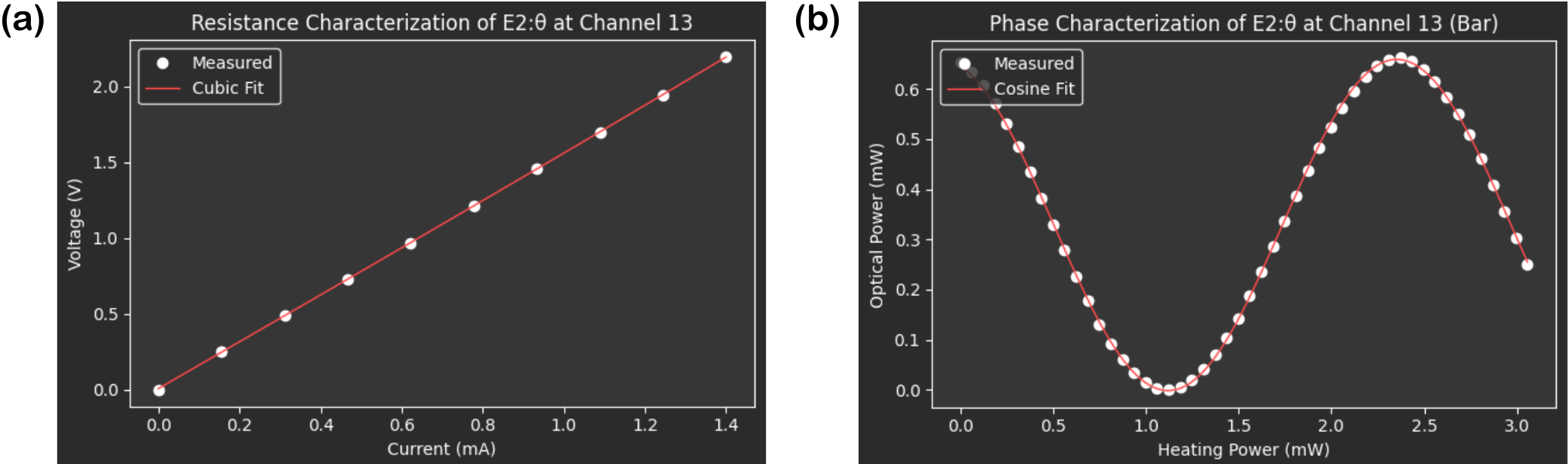}
    \caption{\textbf{Characterization of one thermo-optic phase shifter.}
    (\textbf{a}) Voltage vs.\ current from I-V sweep, showing the cubic fit
    used to extract the nonlinear resistance and thermo-optic efficiency
    coefficient. (\textbf{b}) Optical output power vs.\ applied heating power,
    fitted with a cosine function to extract the static phase offset and the
    current required to realize any target phase.}
    \label{fig:mzi_calibration}
\end{figure}

\subsection*{S1.2 Thermo-optic Phase-Shifter Characterization}

The interferometer mesh comprises 121 thermo-optic phase shifters that were individually characterized using continuous-wave laser light at 1550\,nm and automated Python routines. The internal phase shifters were calibrated using the method described in Ref.~\cite{Alexiev2021}, while the external phase shifters followed the calibration scheme outlined in Ref.~\cite{Lin2024}. In both schemes, each phase shifter is embedded within an on-chip interferometric loop, where it modulates the optical output power. The output from a selected port is measured as a function of electrical heating power, and the resulting interference fringe is fitted to extract the fringe visibility and $\pi$-phase switching power.

The I--V characteristics of the TiN heaters were measured and fitted using a third-order polynomial:
\begin{equation}
V(I) = aI^3 + cI + d,
\end{equation}
where $c$ represents the cold linear resistance and $a$ captures the nonlinear thermal contribution arising from the temperature-dependent resistivity of TiN. The mean cold resistance across all characterized heaters was $1539 \pm 17\,\Omega$. The optical modulation response was modelled using a cosine function of the applied heating power:
\begin{equation}
P_{\mathrm{out}} = A \cdot \cos\!\left(\omega\, P_{\mathrm{heat}} + \phi_0\right) + d,
\end{equation}
where $A$ is the fringe amplitude, $\omega$ is the modulation frequency (inverse of the $2\pi$ switching power), $\phi_0$ is the static phase offset at zero heating power, and $d$ is the vertical offset. The fringe visibility is calculated as $V = A/d$.

Across all 121 characterized phase shifters, we obtained \textbf{an average experimental fringe visibility of $0.9934 \pm 0.0122$ and a mean $2\pi$ modulation period of $2.455 \pm 0.014$\,mW}. This corresponds to a mean $\pi$-phase switching power of $P_\pi = 1.227 \pm 0.007\,\mathrm{mW}$, reflecting the high thermo-optic efficiency of the AMF TiN heater platform. A representative I--V trace and optical modulation curve from one phase shifter are shown in Fig.~\ref{fig:mzi_calibration}, illustrating high fringe contrast and reliable power-to-phase response.

\subsection*{S1.3 Control Electronics}

Thermo-optic phase shifters on the chip are driven using Qontrol Q8iv current driver modules, which provide eight software-defined output channels per module, each delivering current with a precision of 740\,nA. All channels are individually programmable via a custom Python interface, enabling automated, scalable control of the full $12 \times 12$ interferometric mesh.

\vspace{0.5em}
\noindent\textit{More information:} \href{https://qontrol.co.uk/product/q8iv/}{https://qontrol.co.uk/product/q8iv/}

\subsection*{S1.4 Temperature Stabilization}

Temperature control employed two Thorlabs TED200C benchtop TEC controllers, each in conjunction with a thermistor feedback loop. One controller maintained the SPDC crystal at $28.46\,^\circ$C (thermistor resistance setpoint $8.58\,\mathrm{k}\Omega$ on the Thorlabs TH10K calibration curve) to ensure the correct phase-matching condition for wavelength-degenerate photon-pair generation. A second controller maintained the photonic chip at $11.3\,^\circ$C (thermistor resistance setpoint $19\,\mathrm{k}\Omega$), ensuring stable operation, phase coherence, and long-term interferometric contrast throughout all measurements. Both controllers provided temperature stability within $\pm 0.01\,^\circ$C..

\bibliographystyle{naturemag}
\bibliography{sn-bibliography}